# Structural disorder and antiferromagnetism in LaNi$_{1-x}$Pt$_x$O$_3$

A. S. Fjellvåg[a]*, Ø. S. Fjellvåg[b], Y. Breard[c], and A. O. Sjåstad[a]*

[a]Department of Chemistry, Centre for Materials Science and Nanotechnology, University of Oslo, Norway

[b]Department for Neutron Materials Characterization, Institute for Energy Technology, PO Box 40, NO-2027, Kjeller, Norway

[c]Laboratoire CRISMAT, UMR 6508 CNRS ENSICAEN, 6 bd du Maréchal Juin, 14050 Caen Cedex 4, France

*a.o.sjastad@kjemi.uio.no; a.s.fjellvag@smn.uio.no

## 1. Abstract

We report on the *B*-site substitution of Pt in the system LaNi$_{1-x}$Pt$_x$O$_3$. The system can only be synthesized for $x \leq 0.50$, with LaNiO$_3$ ($x = 0.00$) and the stoichiometric double perovskite La$_2$NiPtO$_6$ ($x = 0.50$) as the end members. Higher Pt-contents ($x > 0.50$) are unachievable due to the preference of Pt to either be in oxidation state +IV in octahedral coordination. Upon introducing Pt into LaNiO$_3$, a phase transformation from rhombohedral (*R-3c*) to monoclinic (*P2$_1$/n*) symmetry is observed for $0.075 \leq x \leq 0.125$, where all monoclinic samples are *B*-site ordered, and Pt show a strong preference for the Pt-site. Powder X-ray diffraction analysis reveal disorder of the Pt-distribution in several of the samples with a non-equimolar Ni/Pt ratio ($0.20 \leq x \leq 0.40$), which point toward cluster formation with domains of high and low Pt-content within each sample. La$_2$NiPtO$_6$ further show an antiferromagnetic transition at ~40 K. A similar transition is observed for all monoclinic samples ($x \geq 0.20$), however, the transition becomes weaker for lower *x*. This is explained in light of the structural disorder, *i.e.* by the coexistence of antiferromagnetic domains with long range order and paramagnetic domains dominated by short range antiferromagnetic interactions.

## 2. Introduction

Platinum, one of the most stable noble metals, has been utilized for applications for centuries, e.g. in crucibles for high temperature heat treatments and in thermocouples. This has led to the discovery of several different Pt containing oxides [1], through unexpected side reactions. Pt has preference for oxidation state 0, +II and +IV, as seen from the stability of Pt metal at low partial pressures of oxygen (pO$_2$), Pt$_3$O$_4$ at intermediate temperature and pO$_2$, and PtO$_2$ at higher pO$_2$ [1]. The stability of the +II and +IV states of Pt can be rationalized by the square planar coordination with a d$^8$-configuration (+II) and an octahedral coordination with a d$^6$ low spin configuration (+IV).

In double perovskites, such as La$_2$CoPtO$_6$, Pt adopts the +IV oxidation state. The compound is insulating and weakly ferromagnetic at low temperature (T < ~50 K) [2]. Compared with La$_2$CoIrO$_6$, the magnetic interaction between Co(II)(high spin) and Pt(IV) is weak [2], explained through a hybridization mechanism first applied on Cr-based double perovskites [3]. In these 5d-systems, spin orbit coupling may also affect the magnetic interactions. The synthesis procedure for the similar compound, La$_2$NiPtO$_6$, has previously been reported [4], but its magnetic properties are so far unrevealed. Additionally, neither LaCo$_{1-x}$Pt$_x$O$_3$ nor LaNi$_{1-x}$Pt$_x$O$_3$ have been tried synthesized and investigated for non-equimolar compositions (x ≠ 0.50).

It is well-known that LaNiO$_3$ is a Pauli paramagnetic metal, which upon *B*-site substitution turn insulating. The magnetic properties of LaNiO$_3$ are highly dependent on structural distortions induced by substitution, the choice of substituent element, and the consecutive internal redox processes; the result is varying ferro- (FM) and antiferromagnetic (AFM) interactions [5-10]. In Sr$_2$FeMoO$_6$, the grain structure and degree of *B*-site ordering has been shown to strongly influence both magnetoresistance and magnetic properties [11, 12], also for non-equimolar compositions [13]. By applying various synthesis protocols, different degree of long range Fe-Mo ordering can be achieved. However, the short range Fe-Mo ordering is still present, regardless of the degree of long range ordering. For *B*-site substituted LaNiO$_3$, long range ordering is only achieved by substitution with 4*d*- or 5*d*-elements which obtain a +IV oxidation state, reducing Ni to +II [14]. The subsequent order-disorder phenomena and the effect on magnetic properties in LaNiO$_3$, in both equimolar and non-equimolar compositions, has so far attracted little attention.

Herein we investigate the structure and magnetic properties of LaNi$_{1-x}$Pt$_x$O$_3$, 0 ≤ x ≤ 0.50. Inspired by the studies on Sr$_2$FeMoO$_6$, we pay special attention to the structural disorder of non-equimolar compounds in the system LaNi$_{1-x}$Pt$_x$O$_3$. We further evaluate magnetic properties below room temperature, and correlate low temperature magnetic ordering to structural disorder.

## 3. Experimental

La(NO$_3$)$_3$×6H$_2$O (99.9 %) and Ni(NO$_3$)$_2$×6H$_2$O (99 %) were purchased from Alfa and KEBO Lab, respectively, whereas Pt metal (99.9 %) was purchased from Heraeus and citric acid (C$_6$O$_7$H$_8$×H$_2$O; 98 %) was purchased from Sigma-Aldrich. The La and Ni nitrate salts were dissolved in water and the accurate salt concentrations (mol/g) of the solutions were determined by thermogravimetry. All samples were synthesized using a citric acid complexation method. Stoichiometric amounts of La- and Ni nitrate solutions and Pt metal were weighed out, before the Pt metal was dissolved in aqua regia (1 part HNO$_3$, 3 parts HCl, VWR Chemicals). The two solutions were mixed, and 50 g of citric acid was added per gram of targeted oxide product. The solution was boiled until complete loss of water and nitrate in form of nitrous gasses, followed by overnight heat treatment at 180 °C. The samples were subsequently calcined at 400 °C in air for 12 hours before pelletizing and high temperature heat treatment. For the first annealing step, samples with x ≥ 0.30 were annealed at 1000 °C in sealed quartz ampoules with 7 bar O$_2$ for 12 hours, using Ag$_2$O (KEBO Lab, 99 %) as the oxidizing agent, which releases O$_2$ during the heating sequence whereas samples with x ≲ 0.25 were annealed at 850 °C in a flow of O$_2$ (from AGA, 5.0 quality) for 12 hours. In the second annealing step, all samples were annealed at 1000 °C for 12 hours in a flow of O$_2$. For LaNiO$_3$ (*x* = 0), the high temperature heat treatment was performed twice at 850 °C in a flow of $O_2$ for 48 hours. Additionally, samples with $x = 0.20$, $x = 0.30$ and $x = 0.40$ where annealed again for 2 weeks at 1000 °C in a flow of $O_2$ prior to magnetic measurements, in order to improve the homogeneity of the samples.

Powder synchrotron X-ray diffraction (XRD) data were collected at beam line BM01B at the Swiss-Norwegian Beam Lines (SNBL), European Synchrotron Radiation Facility (ESRF) in Grenoble, France [15]. XRD experiments were performed using a 2D Dexela detector and wavelength $\lambda = 0.50506$ Å. Structural analysis was performed by Rietveld refinements using the software TOPAS [16].

Low temperature synchrotron XRD data were collected at BM01, SNBL at ESRF using a He cryostat (4 to 300 K), $\lambda = 0.71490$ Å using a PILATUS 2M detector. Obtained data were reduced with the Bubble software [17], and the 2D diffraction patterns were visualized with the ALBULA software. All Rietveld refinements were carried out TOPAS [16].

Magnetic measurements were performed using a Quantum Design physical properties measurement system (QD-PPMS). Samples for DC magnetization were measured from 4 to 300 K using a 2000 Oe external DC-field. In AC magnetization mode, samples were measured from 4 to 60 K both with and without a 2000 Oe external DC field, and with a 10 and 15 Oe AC field with frequencies of 117, 1117, 1997, 4997 and 9997 Hz. Field dependent DC magnetization measurements were performed at 4 K with a DC field variating between -9 and 9 Tesla.

## 4. Results

### 4.1 Synthesis and phase relations

In this work, we have investigated $LaNiO_3$, $La_2NiPtO_6$ and $LaNi_{1-x}Pt_xO_3$ for $0 \leq x \leq 0.70$. $LaNiO_3$ is a rhombohedral (*R-3c*) perovskite (Figure 1a), while $La_2NiPtO_6$ is a monoclinic (*P2$_1$/n*) double perovskite with *B*-site ordering (Figure 1b) and a *β*-angle very close to 90°. Attempts to synthesize Pt containing samples by a solid-state route using Pt metal as precursor proved unsuccessful, as the metallic Pt did not oxidize completely and Pt metal impurities were present in the products. As an alternative, we turned to the citric acid complexation synthesis route, where the precursor elements are mixed in solution to ensure good atomic distribution. Combined with a high $pO_2$ during annealing, we avoid that platinum nucleates to Pt metal particles during synthesis. Correspondingly, we find all samples in the range $0 \leq x \leq 0.40$ to be phase pure (Figure S1 and Figure S2). For $x = 0.50$, < 0.50 mol % Pt is present. However, for $x = 0.60$ and $x = 0.70$, $La_2NiPtO_6$ is formed, in addition to significant quantities of Pt and $La_2O_3$ impurities. We believe a higher Pt-content than $x = 0.50$ cannot be stabilized in this system since Pt is unable to adopt lower oxidations states than +IV in the perovskite, and Ni will not go below +II under regular synthesis conditions.

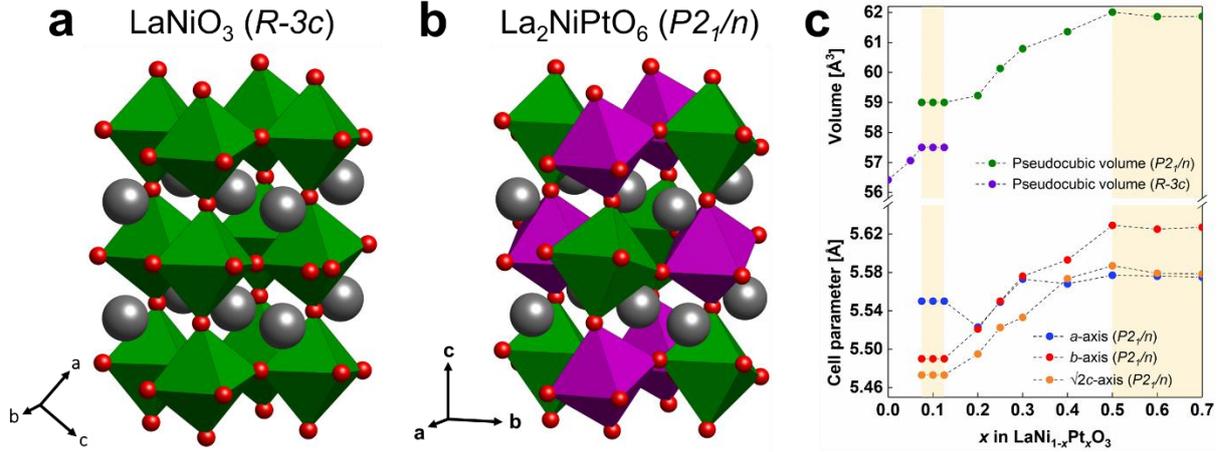

**Figure 1:** Crystal structure of (a) LaNiO$_3$ (*R-3c*) and (b) La$_2$NiPtO$_6$ (*P2$_1$/n*). (c) Cell parameters and pseudocubic cell volume for the LaNi$_{1-x}$Pt$_x$O$_3$ system, where the *c*-axis of the monoclinic phase is multiplied by $\sqrt{2}$ for direct comparison. The rhombohedral lattice parameters (low *x*) are excluded for clarity. The white areas are compositions with single phase, while for the beige areas, multiple phases are present.

Upon the introduction of Pt into LaNiO$_3$, forming LaNi$_{1-x}$Pt$_x$O$_3$, a structural two-phase region is observed for $0.075 \leq x \leq 0.125$ (Figure 1c). This imply that only a small solid solution range is associated with Pt substitution in LaNiO$_3$, while much more Ni-substitution is possible in La$_2$NiPtO$_6$. At the phase boundary, the unit cell volume of the monoclinic phase is ~2.6 % larger than the rhombohedral phase (Figure 1c). Comparing the cell parameters of the monoclinic cell (*c*-axis is multiplied by $\sqrt{2}$), we observe that the three axis lengths become very similar ($a \approx b \approx \sqrt{2}c$) (Figure 1c), *i.e.* there is no strong anisotropic distortion induced by the Pt-substitution. In addition, we note the trend of expansion for the unit cell parameters and volume (Figure 1c) with increasing Pt content (*x*). When entering the regime $x \geq 0.50$, the unit cell parameters and volume flatten out, consistent with unsuccessful introduction of Pt into the compound beyond La$_2$NiPtO$_6$ ($x = 0.50$).

## 4.2 Structural details for LaNi$_{1-x}$Pt$_x$O$_3$ ($0.20 \leq x \leq 0.50$)

The *P2$_1$/n* space group and structure of La$_2$NiPtO$_6$ includes *B*-site ordering [4], while a *B*-site disordered perovskite is better described in the *Pnma* space group. For the higher symmetry perovskite LaNiO$_3$ (*R-3c*), B-site ordering is not allowed by symmetry. In the space group *P2$_1$/n*, *B*-site ordering is indicated by strong intensity of a low angle peak at $Q \approx 1.37$ Å$^{-1}$, which only show intensity if the *B*-site is ordered. Due to the large difference in atomic number between Ni and Pt, we can expect strong intensity of the reflection if the *B*-site is ordered. For the *P2$_1$/n* structure, the peak consists of the (101), (10-1) and (011) Bragg reflections, which all relate to layers with alternating stacking of the two different *B*-site cations (Figure S3). The peak is also allowed in the *Pnma* space group [as (011)], but it always show minimal intensity for disordered perovskites. In this way, the peak is a fingerprint of the long-range ordering of *B*-site cations in perovskites, similar to the descriptions by Meneghini *et al.* [11] on the *B*-site ordering in Sr$_2$FeMoO$_6$.

Rietveld refinements of synchrotron XRD data of $La_2NiPtO_6$ give a good fit using the reported structure model [4] with space group $P2_1/n$ (Figure S2). The cell parameters and volume are very close to the reported values [4]. Refinements of the degree of *B*-site ordering show almost complete B-site ordering for $La_2NiPtO_6$ (Table S1). Ni and Pt are further expected to adopt the oxidation states +II and +IV, respectively, as reported in the literature [2, 4]. This is reasonable considering that the structure is ordered, and in compliance with the empiric rule that *B*-site ordering will only occur if the *B*-elements obtain different oxidation states [14]. Additionally, our magnetic measurements support this conclusion, see below.

Among the monoclinic non-equimolar compositions ($0.20 \leq x \leq 0.40$), the refinements are performed using the same monoclinic structure as $La_2NiPtO_6$ ($P2_1/n$), however, with replacing an amount of Pt with Ni, as in the substitutional system $La_2Ni_{1+y}Pt_{1-y}O_6$ ($0.40 \leq y \leq 1$). For all the monoclinic samples ($0.20 \leq x \leq 0.50$; Figure S2), the analysis provide clear indications of *B*-site ordering, indicated by the intensity of the low angle peak at $Q \approx 1.37$ Å$^{-1}$ (Figure 2). Upon comparing the synchrotron XRD data of the different sample compositions (Figure 2), several trends can be seen with increasing the Pt-content: (1) the ordering-peaks ($Q \approx 1.37$ and 2.64 Å$^{-1}$) intensity increases, (2) all peaks shifts towards lower $Q$, *i.e.* larger interatomic distances, (3) the main perovskite peak ($Q \approx 2.26$ Å$^{-1}$) has a much larger shift towards lower $Q$ compared to the ordering-peaks, and (4) the peak shape for the main perovskite peak changes significantly.

Furthermore, in the Rietveld refinements of the monoclinic samples ($0.20 \leq x \leq 0.50$), the peaks which indicate *B*-site ordering ($Q \approx 1.37$ Å$^{-1}$ and $Q \approx 2.64$ Å$^{-1}$), are not well fitted by the structural model. The two peaks are located at a lower $Q$ than the average structure calculated during Rietveld refinements, *i.e.* they correspond to scattering from a region with a larger unit cell volume than the average structure. A different feature is observed for the main perovskite peak, which for a low Pt-content (low $x$) have an asymmetric tale extending towards higher $Q$, *i.e.* smaller interatomic distances. In this way, it appears as domains of different size contributes differently to the different peaks; *large unit cell domains* contribute to the *B*-site ordering peaks, while *small unit cell domains* contribute to a tail in the main perovskite peak.

Intuitively, the *large unit cell domains* should contain large Pt-contents and simulate the double perovskite $La_2NiPtO_6$, and therefore give a strong scattering contribution to the ordering peaks ($Q \approx 1.37$ and 2.64 Å$^{-1}$). Correspondingly, the *small unit cell domains* may therefore be Ni-rich domains, which contribute to the tail of the main perovskite peak. These factors make it challenging to achieve satisfactory Rietveld refinements with a one structural model. The effect is most prominent for $x = 0.30$, but also strong in $x = 0.20$ and $x = 0.40$, *i.e.* all the monoclinic and non-equimolar compositions.

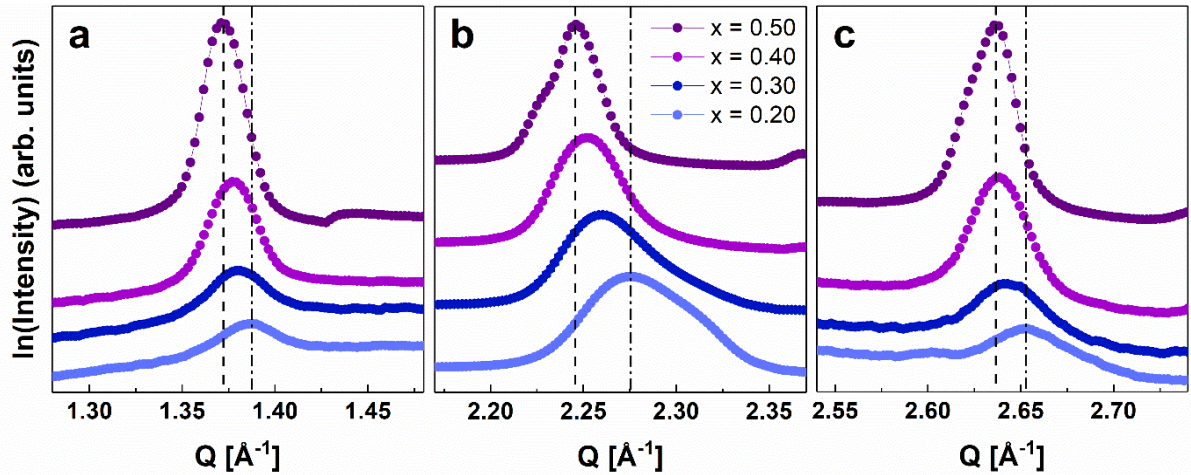

**Figure 2.** Comparison of low angle peak indicating (a) ordering peak with hkl indices (101), (10-1) and (011), (b) main perovskite peak with hkl indices (020), (112) and (200) and (c) a higher angle ordering peak with hkl indices (121), (013), (211) and (103). The compositions are indicated in the figure. The black dashed and dashed dot lines show the position of the maximum of the peaks for $x = 0.50$ and $x = 0.20$, respectively.

To handle this situation, we expand the structural model for Rietveld refinements into a two-phase (domain) model with one *large* and one *small* unit cell (Table S1), mimicking a composite material with different degrees of Ni/Pt content and ordering. In a first approximation, these two phases (domains) were considered to only differ with respect to unit cell parameters. This significantly improved the fit of the asymmetric peak shape and the fit of the position of the ordering peaks, see Figure 3. Subsequent refinements of the *B*-site occupation indicate that the Pt-content is higher in the "large" cell and lower in the "small" cell, consistent with our expectations. The two-phase (domain) model significantly improves the Rietveld refinement fit, which is clearly seen for the main perovskite peak ($Q \approx 2.25$ Å$^{-1}$) when evaluating the difference curve (Figure 3). Though dependable quantification is difficult to obtain, we believe that the qualitative information gathered is reliable. The refinements show that Pt has a very strong preference for the Pt-site, while Ni occupies both sites (Table S1). For samples with a lower Pt-content ($x \leq 0.40$), a lower degree of *B*-site ordering is observed, corresponding to the reduced intensity of the peak at $Q \approx 1.37$ Å$^{-1}$.

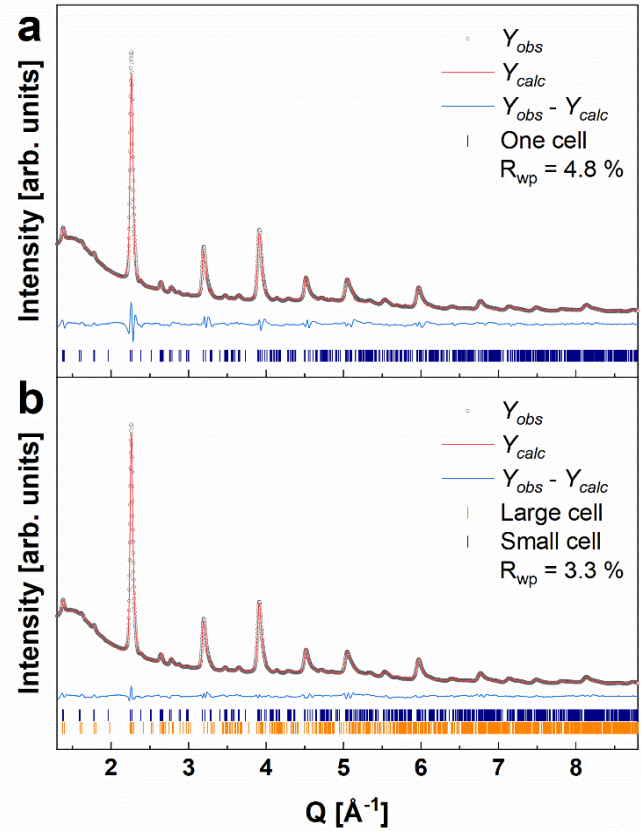

**Figure 3:** Rietveld refinements of LaNi$_{0.70}$Pt$_{0.30}$O$_3$ (*P2$_1$/n*) using (a) a structure model with a single unit cell and using (b) the two-phase (domain) structure model described in the text, which also refine the *B*-site occupancy of Ni and Pt. The fit of the two-phase (domain) model prove superior to reproduce the experimental diffraction pattern.

We believe the "large" and "small" domains originate from inhomogeneous distribution of Pt. Pt-rich domains contain a more ordered Pt-Ni sub-lattice compared to the average structure, while regions poor in Pt are much less *B*-site ordered. The Pt-rich regions are expected to have larger unit cell parameters and volume in view of the trends shown in Figure 1c. The fact that the samples are synthesized from a wet chemical citric acid-based route suggests a homogenous phase with atomic distribution of the elements during phase formation. This is supported by considering *e.g.* the 2D XRD diffraction pattern of $x = 0.30$ (Figure S4) collected at 299 K, which show smooth Debye-Scherrer cones. The homogeneity of the samples imply that the clustering is an intrinsic property of the system, and not a consequence of limited diffusion in the synthesis process.

The Pt-clustering is therefore believed to occur in small nm-sized domains inside grains, similar to the ordering phenomena well described by Meneghini *et al.* for Sr$_2$FeMoO$_6$ [11]. In Sr$_2$FeMoO$_6$, different degree of long-range ordering of Fe and Mo can be achieved by different synthesis protocols. However, the short-range ordering of Fe and Mo is always present, regardless of long range order observed by XRD. For LaNi$_{1-x}$Pt$_x$O$_3$, we may have a very similar situation. The main differences are that (1) Pt has a very high preference for the Pt-site in the structure, and (2) we are additionally considering disorder in non-equimolar compounds.

Based on the current findings, we can now prepare a model on the microstructure of LaNi$_{1-x}$Pt$_x$O$_3$ in the ionic approximation. For La$_2$NiPtO$_6$, a Ni atom located next to a Pt(IV)-octahedra will adopt a +II oxidation state, and the oxygen in between them -II. When a Pt(IV) atom is replaced by a Ni atom, this Ni atom will adopt an oxidation state of +III. To keep charge neutrality, another Ni(III) must also be generated by oxidation of another Ni(II) to Ni(III). Thus, we find it more appropriate to describe the system by the formula La$_2$Ni$_y^{II}$Ni$_{2-2y}^{III}$Pt$_y^{IV}$O$_6$, $0 \leq y \leq 1$, with end members LaNi$^{III}$O$_3$ ($y = 0$) and La$_2$Ni$^{II}$Pt$^{IV}$O$_6$ ($y = 1$). Consequently, Ni(III)-O-Ni(II) segments are present in the structure, and the oxygen atom located between them will be uncompensated in terms of charge. Therefore, there are three different oxygen atoms depending on which elements they are positioned between; Ni(III)-O-Ni(III), Ni(II)-O-Pt(IV) or Ni(II)-O-Ni(III), see Figure 4a. The Ni(II)-O-Ni(III) segment deviate from the ionic approximation and the oxygen atoms of these segments will interact with its nearby Ni-atoms to find the most appropriate charge distribution.

To minimize the energy of the system, the number of uncompensated oxygen atoms must be reduced as much as possible. By evaluating both a clustering of Pt atoms (Figure 4b) and random distribution (Figure 4c), we observe that the number of uncompensated oxygen atoms can be reduced by forming clusters. The Pt rich clusters can in this way form boundaries of uncompensated oxygen atoms, which supports the observed phase separation in the system, simply by an energy minimization. The energy gain for clustering must therefore be significant. On the other hand, if the energy gain was substantial, we would have expected a wider immiscibility gap, with low solubility of Pt into LaNiO$_3$ and of Ni into La$_2$NiPtO$_6$. However, the observed two-phase region is narrow ($0.075 \leq x \leq 0.125$), indicating that the driving force for clustering is strong at a local scale, but not as strong on a long-range scale. The origin behind the phenomena of local ordering and clustering of Pt-rich regions remains to be explained, and local probes such as transmission electron microscopy (TEM), total scattering analysis (PDF) and X-ray absorption spectroscopy (XAS), along with other average structure techniques such as neutron diffraction (ND), may be needed to understand it.

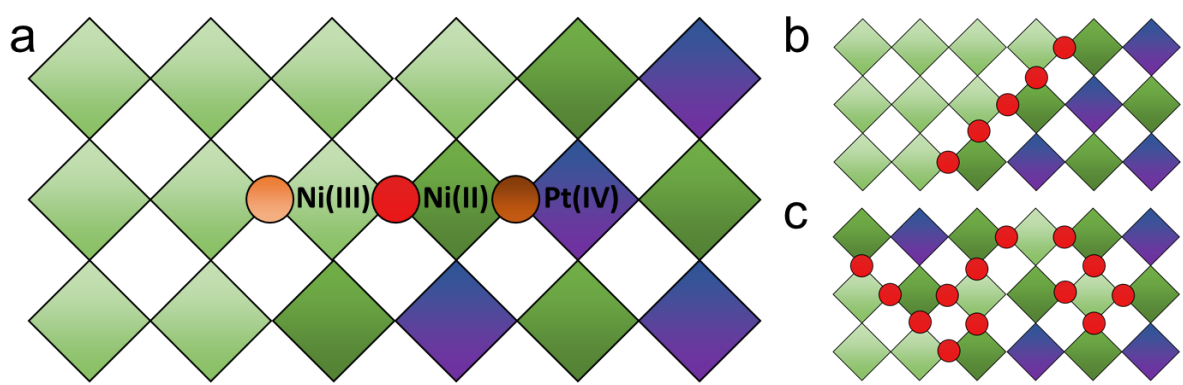

**Figure 4.** (a) Illustration of the different "type" of octahedral corner sharing oxygen atoms (the circles) in LaNi$_{1-x}$Pt$_x$O$_3$. The bright orange oxygen atom is positioned between two Ni(III) atoms, the red is positioned between one Ni(II) and one Ni(III) atom, and the brown in positioned between one Ni(II) and one Pt(IV) atom. (b) Illustration of the number of uncompensated O-atoms in the case of clustering and (c) random Pt-distribution.

## 4.3 Magnetic properties

### 4.3.1 Paramagnetic region (T > 100 K)

LaNiO$_3$ is found to be Pauli paramagnetic in the temperature interval 4-300 K (Figure S5), in compliance with earlier reports on LaNiO$_3$ powders [18]. Upon substituting Pt into LaNiO$_3$, as in LaNi$_{1-x}$Pt$_x$O$_3$, the Pauli paramagnetic behaviour changes to regular temperature dependent paramagnetism for $x = 0.05$, as well for all other samples (Figure 5 and Figure S6). At elevated temperature (~100-300 K), these samples follow a linear Curie-Weiss relationship (Figure S7). For the $x = 0.20$, 0.30 and 0.40 samples prepared, after following only the initial synthesis protocol (2×12 hours' heat treatment), the Curie-Weiss region deviated slightly from linear behaviour. After extended heat treatments (2 additional weeks), the susceptibility is slightly lowered, and the Curie-Weiss region is more pronounced (Figure S8). We suggest this must be related to macro/microstructural changes caused by the heat treatment, despite XRD analysis does not show significant changes after the heat treatment (including the signature ordering peaks) and the average structure is thus assumed to be similar. The peak intensity is generally improved by extended heat treatments, but the peak widths are unchanged.

Curie-Weiss analysis further yields a highly negative $\theta$-value for all analysed compositions, and a paramagnetic moment in the range of 2.4 - 3.1 $\mu_B$, see Table 1. The paramagnetic moment is approximately unchanged through the compositional series, *i.e.* it is almost invariant with respect to the Pt-content. Based on the proposed chemical formula from the ionic approximation (see above), La$_2$Ni$_y^{+II}$Ni$_{2-2y}^{+III}$Pt$_y^{+IV}$O$_6$, and assuming Ni(III) is in the low spin state, the electronic configuration at the *B*-site for $0 \leq x \leq 0.50$ will be:

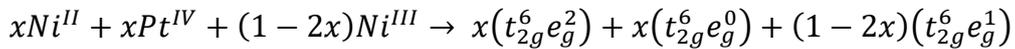

$$xNi^{II} + xPt^{IV} + (1-2x)Ni^{III} \rightarrow x(t_{2g}^6 e_g^2) + x(t_{2g}^6 e_g^0) + (1-2x)(t_{2g}^6 e_g^1)$$

We see that the average number of unpaired electrons at the *B*-site will always be one. In the spin-only approximation [19], this results in a minimal change in the paramagnetic moment through the compositional series (Table 1), and comply well with our observation of a similar paramagnetic magnetic moment for all compositions.

Notably, all compositions show a slightly higher magnetic moment than the spin-only approximation, especially $x = 0.05$, 0.25 and 0.50 (Table 1). The elevated magnetic moment may have several origins, such as small amounts of magnetic impurities, an unexpected high-spin state of Ni(III), spin orbit coupling, or microstructural effects (surfaces-/interfaces). Firstly, magnetic impurities are unlikely. The only impurity observed with XRD is Pt-metal for $x = 0.50$, which does not contribute to the Curie-Weiss paramagnetic moment as Pt is a Pauli paramagnetic metal. The spin state of Ni or spin orbit coupling could in combination explain the elevated magnetic moment, but our experimental data cannot confirm either scenario.

With respect to structural effects, both the degree of *B*-site ordering [12] and the existence of surface states [20] are known to affect magnetic properties. In LaNi$_{1-x}$Pt$_x$O$_3$, the local Ni-Pt ordering inside domains, and the interface between domains, may therefore induce magnetic phenomena not possible to describe by this simple approximation. We have clearly seen the effect of extended heat treatments on the magnetic susceptibility (Table 1 and Figure S8), and believe the sample homogeneity is related to the observed magnetic moment. We therefore conclude that the spin-only approximation describes well the paramagnetic moments of these samples, and that the microstructure contributes with minor additional effects.

Table 1. Experimental magnetic moment and the $\theta$-value calculated using the Curie-Weiss relation, and the theoretical magnetic moment calculated from the spin-only approximation [19].

| $x$ | 0.05 | 0.20* | 0.20** | 0.25 | 0.30 | 0.40 | 0.50 |
|---|---|---|---|---|---|---|---|
| $m$(exp) | 2.7 | 2.7 | 2.4 | 3.1 | 2.6 | 2.4 | 2.9 |
| $m$(theory) | 1.76 | 1.84 | 1.84 | 1.87 | 1.90 | 1.95 | 2.00 |
| $\theta$ [$K$] | -1027 | -350 | -322 | -354 | -194 | -111 | -130 |

*Standard heat treatment procedure, **Sample heated for an additional 2 weeks

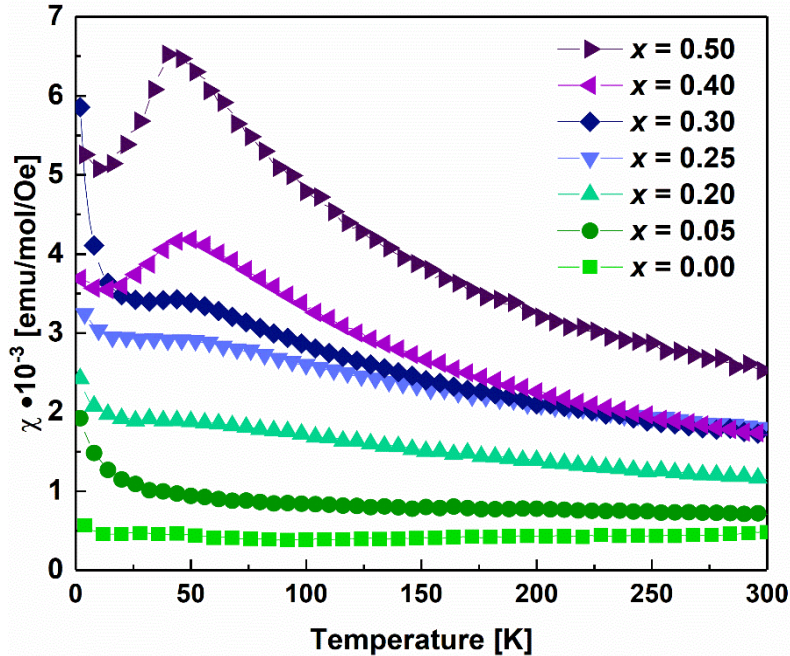

Figure 5. Zero field cooled (ZFC) DC magnetization measurements of LaNi$_{1-x}$Pt$_x$O$_3$ ($0 \leq x \leq 0.50$) using a 2000 Oe external DC-field. The sample compositions are indicated in the figure.

### 4.3.2 Antiferromagnetic transition at low temperature

For La$_2$NiPtO$_6$, the DC magnetic susceptibility show a clear drop below ~40 K, consistent with long-range antiferromagnetic ordering (Figure 5). The transition also occurs for all monoclinic samples ($x \geq 0.20$), *i.e.* with *B*-site ordering, but not for $x = 0.05$. However, the transition gets much less pronounced with reducing Pt-content, and for $x \leq 0.30$ the transition is visible only as a small step in the susceptibility. This is probably due to a significant paramagnetic contribution upon further cooling, resulting from the fact that the non-equimolar samples contain less of the long-range Ni-Pt structural ordering. Still, the transition temperature (T$_N$) is seemingly unaffected and changes very little with Pt-content (Figure 5).

To understand the low temperature magnetic properties of this system, field dependent DC magnetic measurements were performed at 4 K, see Figure 6a. The magnetization is low for all samples, and the lowest magnetization is observed for samples with low $x$ (low Pt-content). It is therefore likely that an AFM ground state dominates all samples, with some differences depending on the Ni/Pt content. This correlates well with the negative theta value from the Curie-Weiss analysis (Table 1), which showed increasingly negative values for samples with a low Pt content (low $x$). Additionally, the field dependent magnetization measurements show an upswing at high magnetic fields for several of the compositions, most prominent for $x = 0.40$ and $x = 0.50$ (Figure 6a). This is evident from the derivative of the magnetization (Figure 6b), which for these samples increases for fields above 1 T. The phenomenon is known as metamagnetism, and the effect is quite week compared to other systems with no signs of saturation up to 9 T.

To further investigate the AFM transition, AC magnetic measurements was performed for two compositions, $x = 0.25$ and $x = 0.50$, see Figure 7. For $La_2NiPtO_6$, $\chi'$ show similar behaviour as the DC magnetization; there is a drop in the susceptibility below ~42 K and the AFM transition show no frequency dependence. $\chi''$ is generally noisy and close to zero, but a weak peak is observed above the transition temperature. This indicates that the AFM transition may start at a higher temperature than indicated from $\chi'$. However, we still believe that $La_2NiPtO_6$ is near being a pure antiferromagnet. For $x = 0.25$, $\chi'$ show an increase in susceptibility below $T_N$ (Figure 7), while $\chi''$ is noisy and close to zero. The behaviour contrasts that of $La_2NiPtO_6$, especially the different behaviour of $\chi'$ (Figure 7), indicating a deviation from pure antiferromagnetism for $x = 0.25$. In this way, all the non-equimolar compositions may be different compared to the equimolar $La_2NiPtO_6$ ($x = 0.50$).

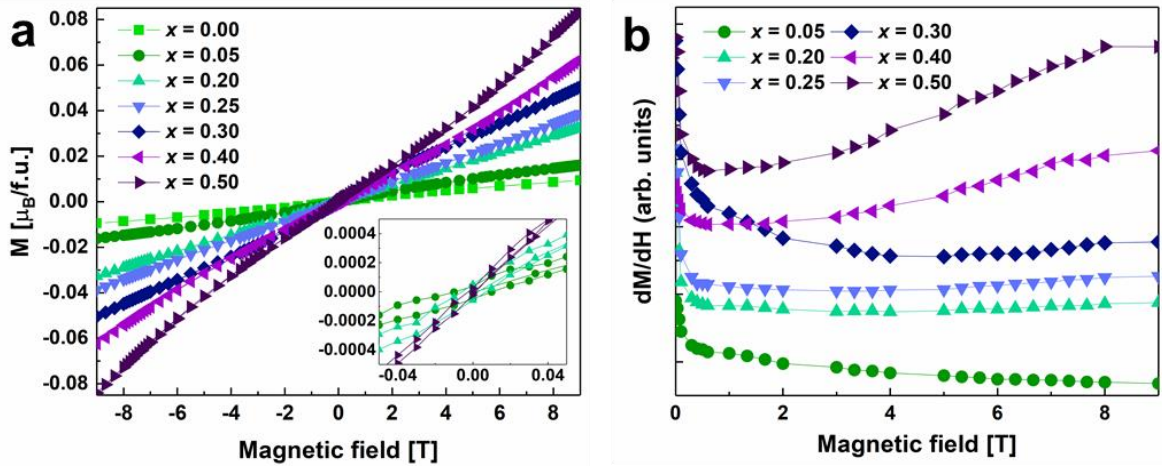

**Figure 6.** (a) Field dependent DC magnetic measurements from -9 to 9 T at 4 K, and (b) the derivative of the magnetization. The formula unit (f.u.) used is $LaNi_{1-x}Pt_xO_3$ for all samples.

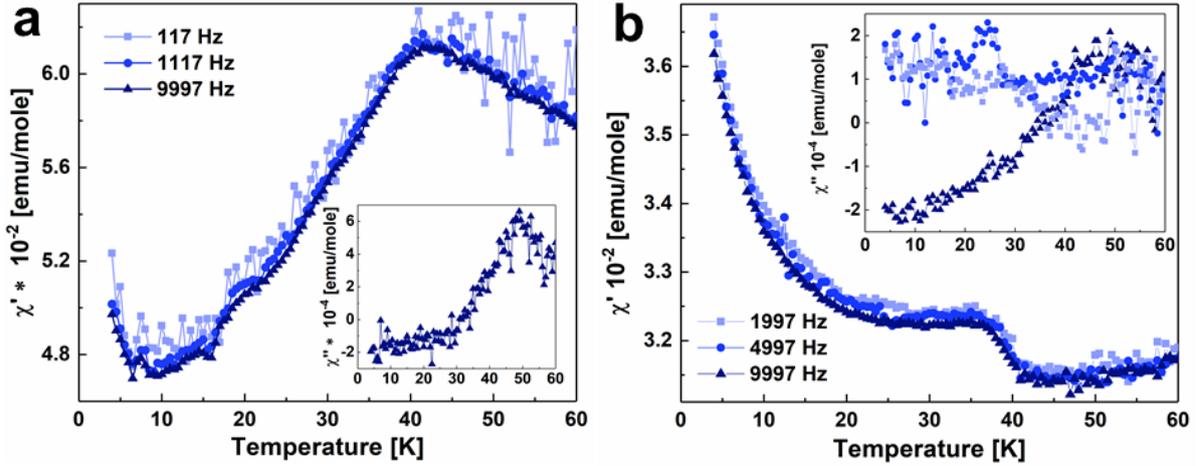

**Figure 7.** AC magnetization measurements of a) La$_2$NiPtO$_6$ using a 10 Oe AC field, and b) of LaNi$_{0.75}$Pt$_{0.25}$O$_3$ using a 15 Oe AC field. The real component of the susceptibility ($\chi'$) is shown in the main figure, and the imaginary component of the susceptibility ($\chi''$) is shown in the inset.

## 5. Discussion

We have shown how Pt-substitution causes disorder in the non-equimolar compositions in LaNi$_{1-x}$Pt$_x$O$_3$ ($0.20 \leq x \leq 0.40$), and that a domain structure appears with regions of high Pt content with large degree of *B*-site ordering, and domains of low Pt-content with low degree of *B*-site ordering. The presented investigation draws similarities to the structural disorder in Sr$_2$FeMoO$_6$ [11]. Locally, Ni-Pt ordering is favoured in a similar way as Fe-Mo ordering is favoured, a property that is highly important for the magnetic interactions in the systems. As presented above, the most evident antiferromagnetism exists in La$_2$NiPtO$_6$, which becomes less pronounced for samples with lower Pt content. However, also the samples with lower Pt-content ($x \leq 0.40$) contain local domains of high Pt-content and *B*-site ordering, which are probably dominated by the same AFM interactions as La$_2$NiPtO$_6$. On the other side, the Pt-depleted domains are most probably dominated by paramagnetism and short range AFM interactions (see highly negative $\theta$-value in Table 1), due to the low Pt-content and missing *B*-site ordering, similar to the sample with $x = 0.05$. This would correlate well with the weak magnetic transition for samples with $x \leq 0.30$ (Figure 5), *i.e.* an AFM transition coexisting with a paramagnetic phase with short range AFM interactions, as shown qualitatively in Figure S9.

The two double perovskites La$_2$NiPtO$_6$ and La$_2$NiTiO$_6$ [21] have several similarities; both are weakly antiferromagnetic at low temperature, and the second *B*-site elements have very similar electronic configurations; t$_{2g}^0$e$_g^0$ for Ti(IV) and t$_{2g}^6$e$_g^0$ for Pt(IV). We speculate if the full t$_{2g}$-orbitals of Pt(IV) are participating, or is only a spectator, with respect to magnetic interactions in La$_2$NiPtO$_6$. If there is no interaction of the full t$_{2g}$-orbitals of Pt, the two compounds should be dominated by similar magnetic interactions, namely the super-super exchange interaction between the e$_g$- and p-orbitals through Ni-O-(Ti/Pt)-O-Ni, as in La$_2$NiTiO$_6$ [21]. However, in La$_2$CoPtO$_6$ [2] a hybridization mechanism is suggested for the interaction between Co(II) high-spin with the full t$_{2g}$-orbital of Pt(IV) [2]. Here, the t$_{2g}$-orbital of Co is only partially filled, allowing the unpaired t$_{2g}$-electrons of Co to hybridize with the Pt t$_{2g}$-orbital. Because the Pt t$_{2g}$- orbital is full, both FM and AFM coupling with Co(II) is possible, resulting in a weak magnetic interaction and low ordering temperature [2]. This shows that even though the $t_{2g}$-orbital of Pt(IV) is full, it can participate in magnetic interactions when the neighbouring element have unpaired $t_{2g}$-electrons. However, in the scenario of Ni(II)($t_{2g}^6 e_g^2$)/Ti(IV)($t_{2g}^0 e_g^2$) and Ni(II)($t_{2g}^6 e_g^2$)/Pt(IV)($t_{2g}^6 e_g^2$), both *B*-site elements have empty/full $t_{2g}$-orbitals, and the $t_{2g}$-orbitals are much less likely to participate in magnetic interactions. It is therefore plausible that the magnetic interactions dominating La$_2$NiPtO$_6$ are more similar to that of La$_2$NiTiO$_6$, with mainly super-super exchange interactions through the $e_g$-p-orbitals, resulting in an AFM ground state for both compounds. We can further expect this AFM interaction to persist for all monoclinic samples in LaNi$_{1-x}$Pt$_x$O$_3$ ($x \geq 0.20$), and it may be the dominating magnetic interaction in the Pt-rich domains with large degree of *B*-site ordering.

The AFM transition is most evident for La$_2$NiPtO$_6$ ($x = 0.50$), but also quite strong for $x = 0.40$. Surprisingly, these two samples are displaying an unexpected metamagnetism (Figure 6). This indicates that the magnetic interaction discussed above, the Ni-O-Pt-O-Ni super-super exchange interaction, is actually quite week, hence the low $T_N$. It is possible that the non-equimolar samples ($0.20 \leq x \leq 0.40$) have short range AFM-interactions that are equally strong or stronger than the super-super exchange interaction, but they lack a specific transition temperature. In the Pt-depleted regions, Ni(III) is the dominating species, which is heavily investigated in both metallic and insulating nickelates [22]. At low temperature, insulating YNiO$_3$ is reported as antiferromagnetic with a charge disproportionation of Ni(III) to Ni(III + $\delta$) and Ni(III − $\delta$), resulting in two different Ni-sites [23]. This disproportionation is also relevant for our compounds, because it shows how Ni(III) is intrinsically unstable in an insulating oxide matrix. Upon Pt-substitution in LaNi$_{1-x}$Pt$_x$O$_3$, the same charge disproportionation of Ni(III) may occur once the compound turn insulating. One can imagine that Ni-atoms located at the Pt-site will exist as Ni(III + $\delta$), while Ni at the Ni-site will exist as Ni(II) when it is near Pt(IV) and Ni(III − $\delta$) when it is near Ni(III + $\delta$). How the electronic landscape of the Ni-rich regions will look is highly difficult to predict, and redox reactions between Ni, Pt and O will probably not result in the perfect ionic picture that is most used in argumentation. Certainly, the local structural arrangement and local magnetic properties will be relevant to understand this system.

## 6. Conclusion

We have demonstrated that the system LaNi$_{1-x}$Pt$_x$O$_3$ ($0 \leq x \leq 0.50$) undergoes a change in oxidation states upon Pt-substitution, causing the system to be better described by the formula La$_2$Ni$_y^{II}$Ni$_{2-2y}^{III}$Pt$_y^{IV}$O$_6$, $0 \leq y \leq 1$. Because Pt has a strong preference for the Pt-site in the monoclinic structure of La$_2$NiPtO$_6$ (*P2$_1$/n*), and obtains only oxidation state +IV, it becomes unfavourable to randomly distribute Pt at the *B*-site due to issues of charge balance on the nearby Ni- and O-atoms. Even a wet-chemical synthesis route results in clustering of Pt-rich and Pt-depleted domains inside the sample, demonstrating that it is an energetically favourable state. All samples show antiferromagnetic interactions (highly negative $\theta$-value), while only the Pt-rich samples with a large degree of structural long-range order turn to a clear antiferromagnetic state at low temperature ($T_N \approx 40$ K). This is most probably due to a similar antiferromagnetic interaction as in $La_2NiTiO_6$ (super-super-exchange). The strength of this magnetic interaction is uncertain, as a weak metamagnetism is observed for $x = 0.40$ and $x = 0.50$, but not for the other compositions.

## 7. Acknowledgements

The authors would like to acknowledge the help of Dr. Susmit Kumar with magnetic measurements and discussions on this topic. We also acknowledge the expertise of the Swiss-Norwegian Beam Lines at ESRF, Grenoble. In addition, we greatly appreciate discussions on structure-property relations in the NAFUMA group at the University of Oslo. The project was financed by the Research Council of Norway through the projects RIDSEM (project no. 272253) and iCSI (project no. 237922).

## 8. Supplementary

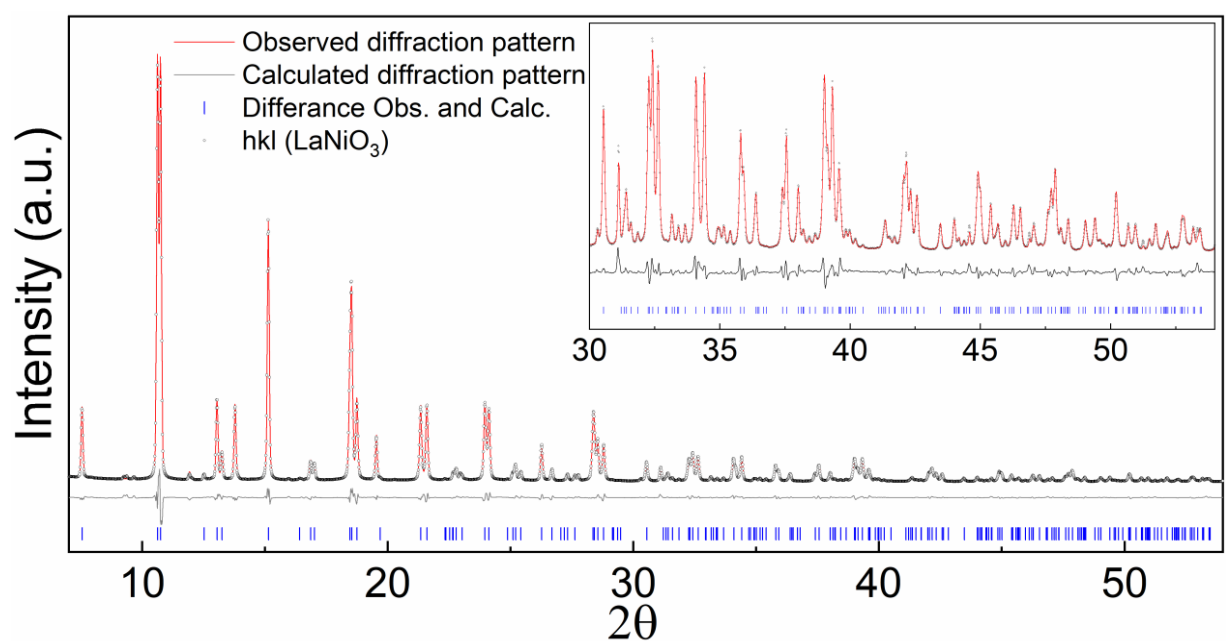

**Figure S1. Rietveld refinements of $LaNiO_3$. The refinements were performed using space group *R-3c*; $\lambda = 0.50506$ Å.**

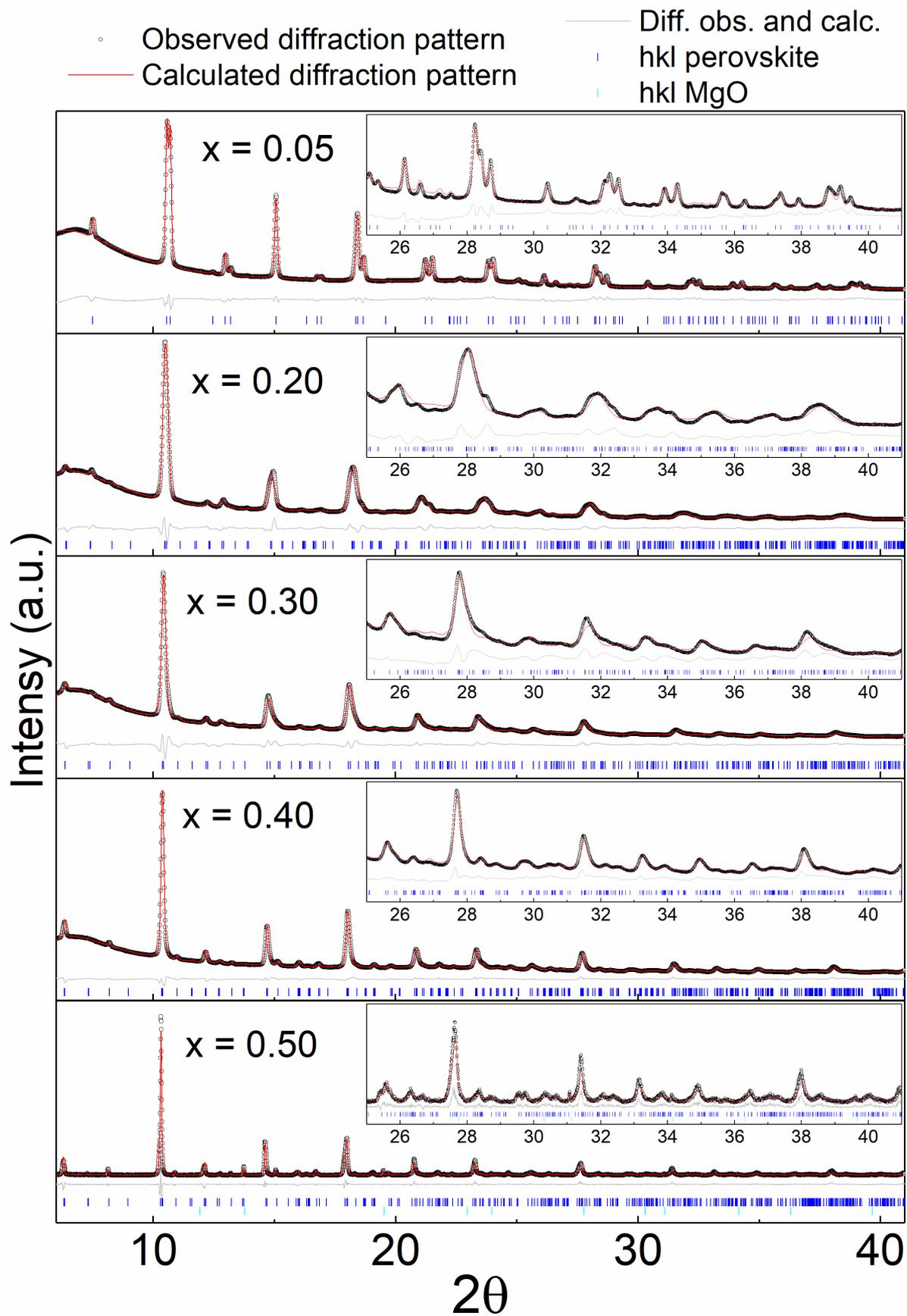

**Figure S2.** Rietveld refinement of **LaNi$_{1-x}$Pt$_x$O$_3$**. The refinements were performed using space group *R-3c* for the sample with $x = 0.05$ and space group *P2$_1$/n* for the samples with $x \geq 0.20$; $\lambda = 0.50506$ Å.

**Table S1. Results from structural analysis and Rietveld refinements of selected members in the LaNi$_{1-x}$Pt$_x$O$_3$ series.**

| x, domain model | Unit cell size | Phase fraction | a [Å] | b [Å] | c [Å] | beta [°] | Volume [Å$^3$] | Rwp | Cry. Size [nm] | Pt-occ (theory) | Pt-occ (exp) |
|---|---|---|---|---|---|---|---|---|---|---|---|
| 0.05 | "large" | 0.55 | 5.485 |  | 13.181 |  | 343.4 |  | 39 | 0.05 | 0.12 |
| 0.05 | "small" | 0.45 | 5.470 |  | 13.155 |  | 340.9 | 3.25 | 73 | 0.05 | 0.04 |
| 0.20 | "large" | 0.59 | 5.574 | 5.528 | 7.743 | 90.64 | 238.6 |  | 31 | 0.40 | 0.43 |
| 0.20 | "small" | 0.41 | 5.500 | 5.500 | 7.751 | 91.67 | 234.3 | 3.43 | 91 | 0.40 | 0.21 |
| 0.30 | "large" | 0.7 | 5.577 | 5.586 | 7.818 | 90.47 | 243.6 |  | 115 | 0.60 | 0.63 |
| 0.30 | "small" | 0.3 | 5.584 | 5.516 | 7.731 | 91.41 | 238.0 | 3.33 | 22 | 0.60 | 0.33 |
| 0.40 | "large" | 0.87 | 5.582 | 5.604 | 7.865 | 90.30 | 246.0 |  | 55 | 0.80 | 0.82 |
| 0.40 | "small" | 0.13 | 5.603 | 5.560 | 7.785 | 91.14 | 242.4 | 3.17 | 36 | 0.80 | 0.35 |
| 0.50 | "large" | 0.54 | 5.571 | 5.638 | 7.904 | 90.12 | 248.3 |  | 116 | 1.00 | 1.00 |
| 0.50 | "small" | 0.46 | 5.589 | 5.618 | 7.895 | 90.13 | 247.9 | 2.96 | 139 | 1.00 | 0.97 |
| x, simple model | Unit cell size | Phase fraction | a [Å] | b [Å] | c [Å] | beta [°] | Volume [Å$^3$] | Rwp | Cry. Size [nm] | Pt-occ (theory) | Pt-occ (exp) |
| 0.05 | avg | 1 | 5.479 |  | 13.170 |  | 342.4 | 4.13 | 115 | 0.050 |  |
| 0.20 | avg | 1 | 5.545 | 5.538 | 7.751 | 91.04 | 238.0 | 3.64 | 107 | 0.400 | 0.3789 |
| 0.30 | avg | 1 | 5.573 | 5.587 | 7.809 | 90.75 | 243.1 | 4.76 | 107 | 0.600 | 0.5765 |
| 0.40 | avg | 1 | 5.584 | 5.602 | 7.866 | 90.36 | 246.0 | 4.59 | 107 | 0.800 | 0.7839 |
| 0.50 | avg | 1 | 5.578 | 5.629 | 7.902 | 90.12 | 248.2 | 3.35 | 135 | 1.000 | 0.9679 |

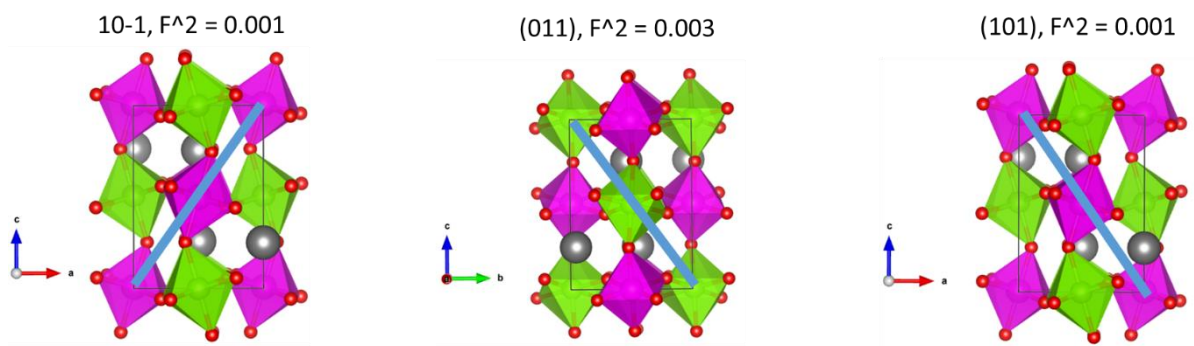

**Figure S3.** Example of crystallographic planes associated with *B*-site ordering in the double perovskite La$_2$NiPtO$_6$ (*P2$_1$/n*).

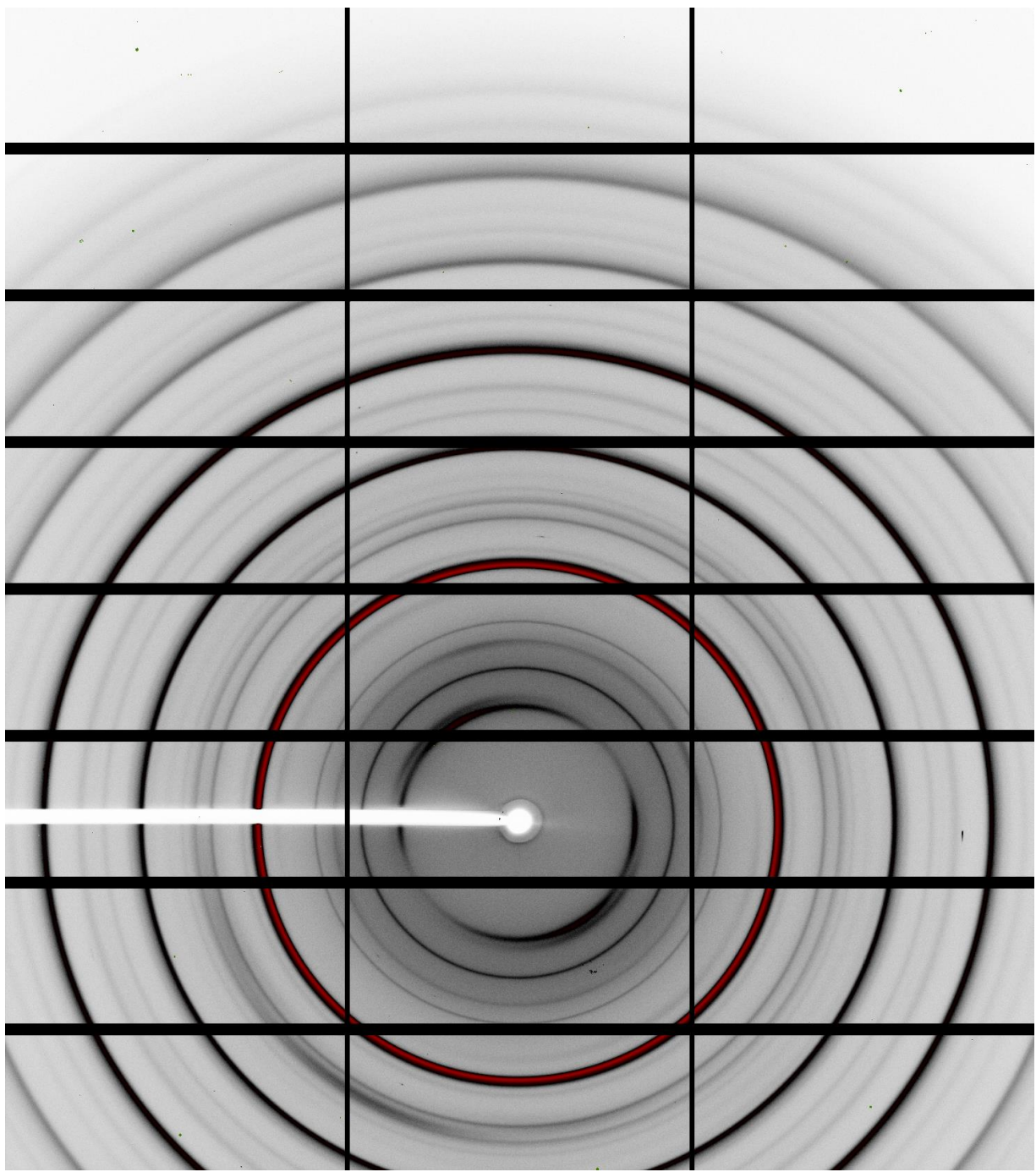

**Figure S4.** X-ray diffraction pattern from 2D-detector of the sample LaNi$_{0.70}$Pt$_{0.30}$O$_3$ collected at 299 K. The variating intensity in the inner rings is scattering from the sample setup. The picture is generated with the ALBULA software.

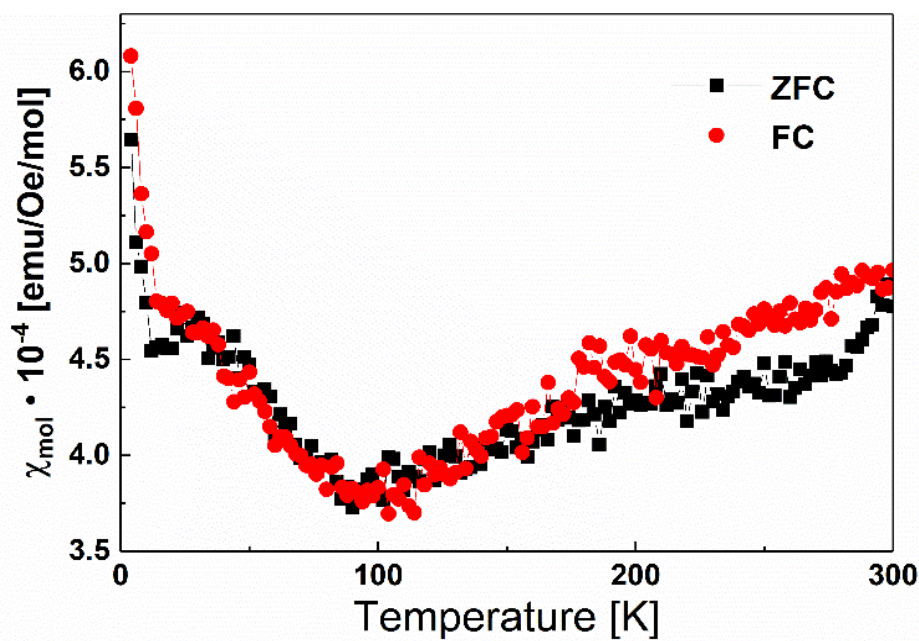

**Figure S5. ZFC and FC magnetic measurement of LaNiO$_3$.**

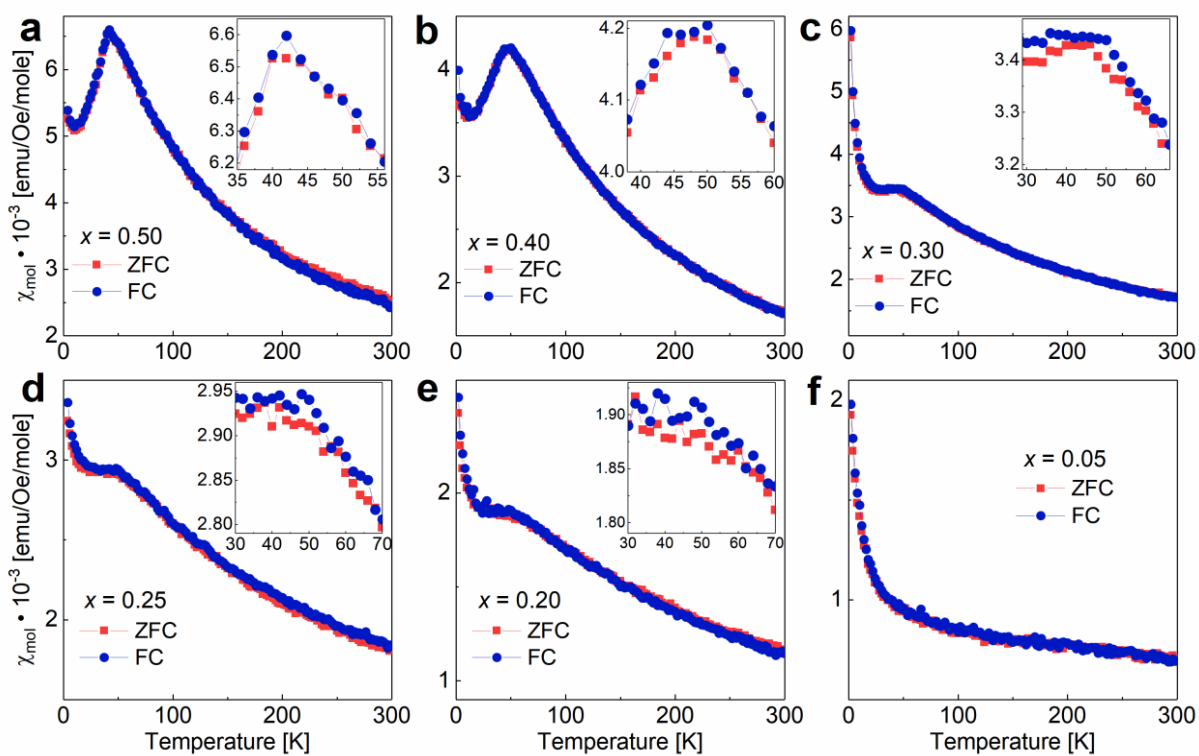

**Figure S6. Zero field cooled (ZFC) and field cooled (FC) DC magnetization measurements of LaNi$_{1-x}$Pt$_x$O$_3$ ($0 \leq x \leq 0.5$) using a 2000 Oe external DC-field. The sample compositions are indicated in the figure.**

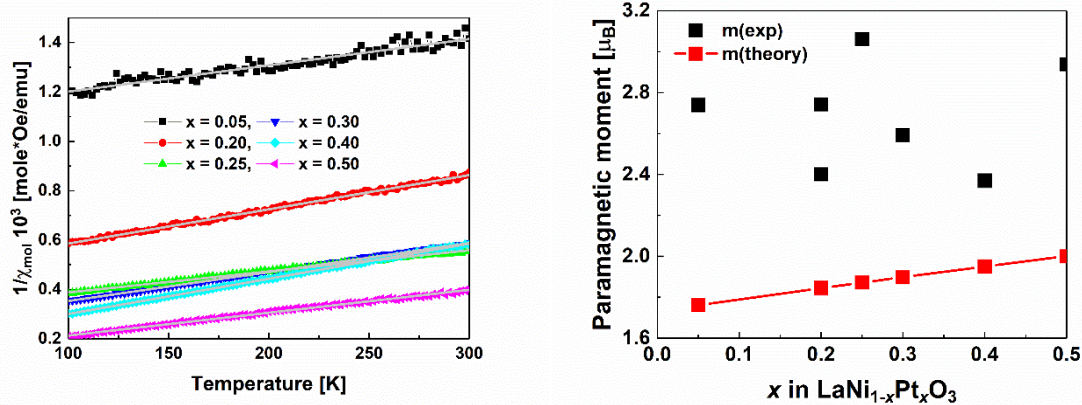

**Figure S7.** Linear fit using the Curie-Weiss relation (left) and the obtained magnetic moment compared with theoretical magnetic moment (right).

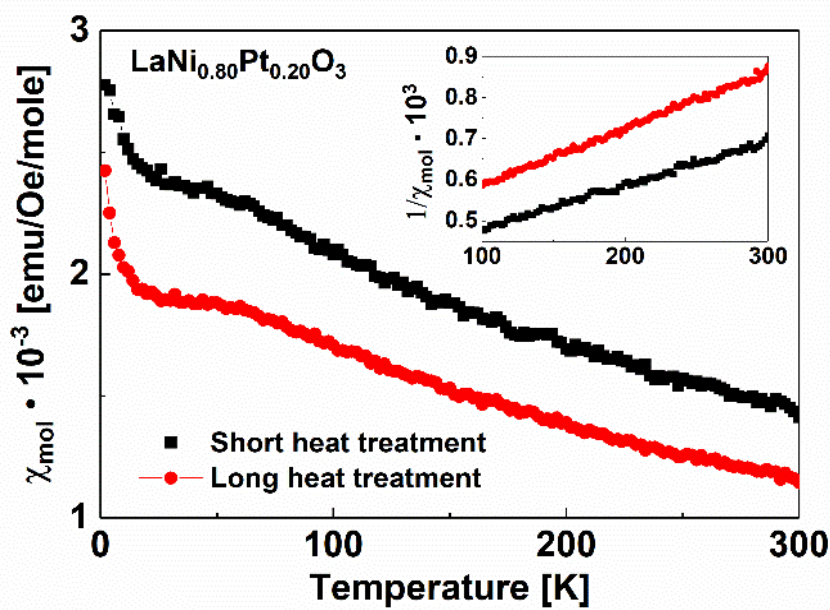

**Figure S8.** Comparison between ZFC magnetic measurement of LaNi$_{0.80}$Pt$_{0.20}$O$_3$ after a short (2×12 hours) and long (additional 2 weeks) heat treatment. $\chi_{mol}$ is shown in the figure and $1/\chi_{mol}$ is shown in the inset.

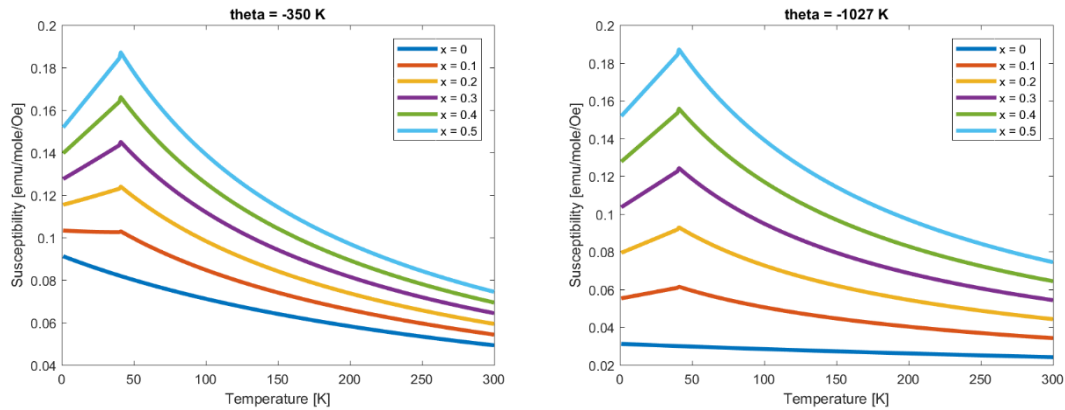

**Figure S9.** Example on susceptibility of a combination of an antiferromagnetic and a paramagnetic phase coexisting in a sample. The value of $x$ represents the substitution, where $x = 0$ corresponds to a paramagnetic samples and x = 0.5 corresponds to a AFM samples, as in $LaNi_{1-x}Pt_xO_3$. The data is calculated from the Curie-Weiss law for the paramagnetic phase and for the AFM phase above $T_N = 40$ K, and uses a linear decline in susceptibility for the AFM phase below $T_N$. Both the paramagnetic and AFM phase use a m = 2 μB. For the AFM phase, θ = –130 K, while for the paramagnetic phase, θ = –350 K (left) and θ = –1027 K (right). This corresponds to the data in Table 1 for the samples $x = 0.50$, $x = 0.20$ and $x = 0.05$, respectively.